# Non-Relativistic Particle under the Influence of Aharonov-Bohm Flux Field Subject to Physical Potentials and the Nikiforov-Uvarov Method

FAIZUDDIN AHMED*, KAYSER AHMED, AKHERUZZAMAN AHMED, ARIFUL ISLAM, BIKASH PRATIM BARMAN

*Department of Physics, University of Science & Technology Meghalaya, Ri-Bhoi, Meghalaya-793101, India*

(*Corresponding author: [faizuddinahmed15@gmail.com](mailto:faizuddinahmed15@gmail.com))

## Abstract

In this work, the non-relativistic wave equation via the Schrödinger wave equation under the influence of the Aharonov-Bohm flux field Subject to physical potentials of various kinds is investigated. These potentials are modified Coulomb potential, modified harmonic oscillator potential, the Kratzer-Feus potential, and the Mie-type potential which have wide applications in different branches of physics and chemistry. We solve the Schrodinger wave equation using the Nikiforov-Uvarov (NU) method and obtain the energy profiles and the wave-function of the non-relativistic particle, and analyze the effects of potential and the quantum flux on them. We show that each non-relativistic energy level gets modified in comparison to the known results obtained in the literature.

**Keywords**: Schrodinger wave equation, Nikiforov-Uvarov method, solutions of wave equation: bound-states, interaction potential, geometric quantum phase

# 1. Introduction:

The Schrodinger equation is a linear partial differential equation that governs the wave function of a quantum mechanical system. It is the key result in quantum mechanical system and its discovery was a significant landmark in the development of the subject. The time-dependent Schrödinger equation is represented by [1--4]

$$i \hbar \frac{d\Psi}{dt} = -\frac{\hbar^2}{2M}\nabla^2\Psi + V\Psi. \qquad (1)$$

The time-independent non-relativistic wave equation is given by

$$\left[\nabla^2 + \frac{2M}{\hbar^2}(E - V(r))\right]\Psi = 0. \qquad (2)$$

Here, $\nabla^2$ is the Laplacian operator in a coordinate system, $\Psi$ is the wave fuvction that assigns a complex number. The parameter M is the mass of the particle, and V(r) is the potential that represents the environment in which the particle exists.

Over several decades, there has been a growing interest among researchers to obtain the analytical solutions of the Schrodinger equation for physical potential in quantum mechanical systems. The exact solution of the Schrödinger equation (SE) with spherically symmetric potential plays a vital role in different branches of physics including nuclear physics, atomic and molecular physics, and in modern physics. These potentials play important roles in different fields of physics such as; plasma, solid state and atomic physics [5]. Some of these potentials include the Cornell potential [6,7], mixed between the Cornell potential and the harmonic oscillator potential [8,9],generalized Morse potential [10], Yukawa potential [11], Wood-Saxon potential [12], Hulthen potential [13], Eckart potential [14], Makorov potential [15], Hellmann potential [16], Coulomb potential [17], the harmonic oscillator potential [18],

Pseudo-harmonic potential [19], Mie-type potential [20], quark-antiquark interaction potential [21].

Several Researchers have been solved the Schrödinger equation using different methods [22—26], which includes asymptotic iteration method (AIM) [27—33], the super-symmetric shape invariance method [34—39], the Nikiforov-Uvarov (NU) method [24, 40—47], the variational method [48]. For instance, Ita [49] solved the Schrödinger equation with the Hellmann potential and obtained the energy eigenvalues and the corresponding wave functions using both 1/N expansion method and the NU method. Kocak et al. [50] in their work solved the Schrödinger equation with the Hellmann potential using the asymptotic iteration method and obtained energy eigenvalues and the wave functions. Mesa et al. [51] in their study, obtained bound state spectrum of the Schrödinger equation with generalized Morse potential and Poschle-Teller potential respectively. Arda *et al.* [52] solved one-dimensional Schrödinger equation for the generalized Morse potential with the NU method and obtained the bound state solutions for the effective mass for some diatomic molecules. Similarly, Zhang *et al.* [11] solved any state solutions of the Schrödinger equation with the generalized Morse potential using the basic concept of the super symmetric shape invariance formalism and the function analysis method. It is important to note that the solutions of a combination of two or more of these potentials give a significant result with diverse applications. For instance, Onate *et al.* [53] obtained the solutions of the radial Schrödinger equation with a combination of Coulomb potential, Yukawa potential and inversely quadratic Yukawa potential (class of Yukawa potentials) which has application in plasma physics, solid state physics and atomic physics. Awoga *et al.* [54] obtained the solutions of the Schrödinger equation with generalized inverted hyperbolic potential. Onate *et al.* [55-56], obtained analytical solutions of both Dirac equation and Klein Gordon equation with Hellmanne-Froste-Musulin potential (combination of Hellmann potential and Froste-Musulin potential) and combined potential (combination of general Manninge-Rosen potential, hyperbolical potential and Pöschle-Teller potentials). The HellmanneFrosteMusulin potential has application in condense matter physics, atomic and molecular physics while the combined potential has application in high energy physics, nuclear physics, atomic and molecular physics. Since the combination of two or more potentials gives better result especially when a stabilizing potential like the Yukawa is added, stability of the nucleus is achieved and has been used extensively in nuclear physics [57---61]. For instance, Hamzavi *et al.* [58] applied the inversely quadratic Yukawa potential and a tensor interaction term to the solutions of the approximate spin and pseudo-spin symmetries of the Dirac. Their results show that by applying the tensor interaction term, the degeneracy between spin and pseudo-spin state doublets were removed. The harmonic oscillator potential is used in many branches of physics.

In this paper we have used few modified known potentials in literature as follows:

(a) Modified Coulomb potential

$$V(r) = a - \frac{b}{r} \tag{3}$$

(b) Modified harmonic oscillator potential

$$V(r) = a + b\, r^2 \tag{4}$$

(c) The Kratzer-Feus potential (Inverse square plus Coulomb potential)

$$V(r) = -\frac{b}{r} + \frac{c}{r^2} \tag{5}$$

(d) The Mie-type Potential (constant plus inverse square and Coulomb potential)

$$V(r) = a - \frac{b}{r} + \frac{c}{r^2} \tag{6}$$

We solve the non-relativistic wave equation under the effects of above mentioned potentials in the presence of Aharonov-Bohm flux field with the help of the Nikiforov-Uvarov method. We show that the presence of quantum flux modified the energy level and wave function of a Schrodinger particle.

## 2. Brief review of the Nikiforov-Uvarov method:

The Nikifrove-Uvarov metho is helpful in order to find eigenvalue and eigenfunctions of the Schrödinger equation, as well as other second-order differential equations of physical interest. This NU method has been successfully applied in solving the quantum mechanical problems by several researchers [24, 40—47, 62—68].

According to this method, eigenfunction second-order differential equations [69—71]

$$\frac{d^2\Psi(s)}{ds^2} + \frac{(\alpha_1 - \alpha_2 s)}{s(1-\alpha_3 s)}\frac{d\Psi(s)}{ds} + \frac{(-\xi_1 s^2 + \xi_1 s - \xi_3)}{s^2(1-\alpha_3 s)^2}\Psi(s) = 0 \quad (7)$$

are given by

$$\Psi(s) = s^{\alpha_{12}}(1-\alpha_3 s)^{-\alpha_{12}-\frac{\alpha_{13}}{\alpha_3}} P_n^{\left(\alpha_{10}-1,\frac{\alpha_{11}}{\alpha_3}-\alpha_3-1\right)}(1-2\alpha_3 s) \quad (8)$$

And that the eigen energy eigen values equations

$$\alpha_2 n - (2n+1)\alpha_2 + (2n+1)(\sqrt{\alpha_9} + \alpha_3\sqrt{\alpha_8}) + n(n-1)\alpha_3 + \alpha_7 + 2\alpha_3\alpha_8 + 2\sqrt{\alpha_8\alpha_9} = 0 \quad (9)$$

The parameters $\alpha_4 \ldots \alpha_{13}$ are obtained from the six parameters $\alpha_1 \ldots \alpha_3$ and $\xi_1 \ldots \xi_3$ as follows:

$$\alpha_4 = \frac{1}{2}(1-\alpha),$$

$$\alpha_5 = \frac{1}{2}(\alpha_2 - 2\alpha_3),$$

$$\alpha_6 = \alpha_5^2 + \xi_1,$$

$$\alpha_7 = 2\alpha_4\alpha_5 - \xi_2,$$

$$\alpha_8 = \alpha_4^2 + \xi_3,$$

$$\alpha_9 = \alpha_6 + \alpha_3\alpha_7 + \alpha_3^2\alpha_8$$

$$\alpha_{10} = \alpha_1 + 2\alpha_4 + 2\sqrt{\alpha_8},$$

$$\alpha_{11} = \alpha_2 - 2\alpha_5 + 2(\sqrt{\alpha_9} + \alpha_3\sqrt{\alpha_8}),$$

$$\alpha_{12} = \alpha_4 + \sqrt{\alpha_8},$$

$$\alpha_{13} = \alpha_5 - (\sqrt{\alpha_9} + \alpha_3\sqrt{\alpha_8}). \tag{10}$$

A special case where $\alpha_3 = 0$, we find

$$lim_{\alpha_3 \to 0} P_n^{\left(\alpha_{10}-1, \frac{\alpha_{11}}{\alpha_3}-\alpha_{10}-1\right)}(1 - 2\alpha_3 s) = L_n^{\alpha_{10}-1}(\alpha_{11}s), \text{ and}$$

$$\Rightarrow lim_{\alpha_3 \to 0}(1 - \alpha_3 s)^{-\alpha_{12}-\frac{\alpha_{13}}{\alpha_3}} = e^{\alpha_{13}s}. \tag{11}$$

Therefore the wave-function from (A.2) becomes

$$\Psi(s) = s^{\alpha_{12}} e^{\alpha_{13}s} L_n^{\alpha_{10}-1}(\alpha_{11}s), \tag{12}$$

Where $L_n^{(\beta)}(x)$ denotes the generalized Liguerre polynomial.

The energy eigenvalues equation reduces to

$$n\alpha_2 - (2n+1)\alpha_5 + (2n+1)\sqrt{\alpha_9} + \alpha_7 + 2\sqrt{\alpha_8\alpha_9} = 0 \tag{13}$$

## 3. MOTION OF NON-RELATIVISTIC PARTICLE UNDER SPHERICALLY SYMMETRIC POTENTIAL IN THE PRESENCE OF AB-FLUX FIELD:

The time independent Schrödinger equation in spherical system is described by the Eq. (2) given by

$$\nabla^2 \Psi(r, \theta, \phi) + \frac{2M}{\hbar^2}(E - V(r)) \Psi(r, \theta, \phi) = 0. \tag{14}$$

As potential is spherically symmetric, it is convenient work in spherical polar coordinate, (r, θ, ϕ) where r ≤ θ≤ ∞,0≤ θ ≤ π, 0≤ ϕ≤ 2π. Expressing Laplacian operator in spherical coordinates, we have

$$\left[\frac{1}{r^2}\frac{d}{dr}\left(r^2\frac{d}{dr}\right) + \frac{1}{r^2\sin\theta}\frac{d}{d\theta}\left(sin\theta\frac{d}{d\theta}\right) + \frac{1}{r^2 sin^2\theta}\frac{d^2}{d\phi^2} + \frac{2M}{\hbar^2}\left(E - V(r)\right)\right]\Psi(r,\theta,\phi) = 0. \qquad (15)$$

**Separation of the equation:**

The wave function of Eq. (15) can be written as

$$\Psi(r, \theta, \phi) = R(r)\,\Theta(\theta)\,\Phi(\phi). \qquad (16)$$

Substituting this form of the wave-function ψ into the Eq. (15) and multiplying by $\frac{r^2 sin^2\theta}{R\Theta\phi}$, we have obtained

$$\frac{sin^2\theta}{R}\frac{d}{dr}\left(r^2\frac{dR}{dr}\right) + \frac{sin\theta}{\Theta}\frac{d}{d\theta}\left(sin\theta\frac{d\Theta}{d\theta}\right) + \frac{2M}{\hbar^2}[E - V(r)]r^2 sin^2\theta = -\frac{1}{\Phi}\frac{d^2\Phi}{d\phi^2}. \qquad (17)$$

The left hand side of Eq. (4) is a function of r and θ and right side is a function of ϕ alone. This is possible when each side is a constant, say $m^2$. Then

$$\frac{d^2\Phi}{d\phi^2} = -m^2\Phi. \qquad (18)$$

And

$$\frac{sin^2\theta}{R}\frac{d}{dr}\left(r^2\frac{dR}{dr}\right) + \frac{sin\theta}{\Theta}\frac{d}{d\theta}\left(sin\theta\frac{d\Theta}{d\theta}\right) + \frac{2M}{\hbar^2}[E - V(r)]r^2 sin^2\theta = m^2. \qquad (19)$$

Dividing both sides of Eq. (19) by $sin^2\theta$ and renaming we get,

$$\frac{1}{R}\frac{d}{dr}\left(r^2\frac{dR}{dr}\right) + \frac{2M}{\hbar^2}(E-V)r^2 = -\left[\frac{1}{\Theta sin\theta}\frac{d}{d\theta}\left(sin\theta\frac{d\Theta}{d\theta}\right) - \frac{m^2}{sin^2\theta}\right]. \qquad (20)$$

This is possible when both sides are equal to a constant, λ. Consequently, we get the Θ- and the radial R-equation as follows:

$$\frac{1}{\sin\theta}\frac{d}{d\theta}\left(\sin\theta\frac{d\Theta}{d\theta}\right) + \left(\lambda - \frac{m^2}{\sin^2\theta}\right)\Theta = 0. \quad (21)$$

And

$$\frac{1}{r^2}\frac{d}{dr}\left(r^2\frac{dR}{dr}\right) + \frac{2M}{\hbar^2}[E - V(r)]R - \frac{\lambda}{r^2}R = 0$$

$$\Rightarrow \frac{d^2R}{dr^2} + \frac{2}{r}\frac{dR}{dr} + \frac{2M}{\hbar^2}\left[(E - V(r)) - \frac{\lambda\hbar^2}{2Mr^2}\right]R = 0. \quad (22)$$

Substituting $R = \frac{U(r)}{r^{\frac{1}{2}}}$ into the Eq. (22), we have

$$\Rightarrow U''r^{-\frac{1}{2}} - U'r^{-\frac{3}{2}} + \frac{3}{4}Ur^{-\frac{5}{2}} + \frac{2}{r}\left[-\frac{1}{2}Ur^{-\frac{3}{2}} + Ur^{-\frac{1}{2}}\right] + \frac{2M}{\hbar^2}\left[E - V(r) - \frac{\lambda\hbar^2}{2Mr^2}\right]Ur^{-\frac{1}{2}}$$

$$= 0$$

$$\Rightarrow U''r^{-\frac{1}{2}} - U'r^{-\frac{3}{2}} + \frac{3}{4}Ur^{-\frac{5}{2}} - Ur^{-\frac{5}{2}} + 2Ur^{-\frac{3}{2}} + \frac{2m}{\hbar^2}\left[E - V - \frac{\lambda\hbar^2}{2mr^2}\right]Ur^{-\frac{1}{2}} = 0$$

$$\Rightarrow U''r^{-\frac{1}{2}} + U'r^{-\frac{3}{2}} - \frac{1}{4}Ur^{-\frac{5}{2}} + \frac{2M}{\hbar^2}\left[E - V - \frac{\lambda\hbar^2}{2Mr^2}\right]Ur^{-\frac{1}{2}} = 0$$

$$\Rightarrow U'' + U'r^{-1} - \frac{1}{4}Ur^{-2} + \frac{2M}{\hbar^2}\left[E - V - \frac{\lambda\hbar^2}{2Mr^2}\right]U = 0$$

$$\Rightarrow U'' + \frac{1}{r}U' + \left[\frac{2M}{\hbar^2}(E - V) - \frac{1}{r^2}\left(\lambda + \frac{1}{4}\right)\right]U = 0$$

$$\Rightarrow U'' + \frac{1}{r}U' + \left[\frac{2M}{\hbar^2}(E - V) - \frac{J^2}{r^2}\right]U = 0. \quad (23)$$

Where, we have written $\lambda = l(l+1)$ and $J = \sqrt{\lambda + \frac{1}{4}} = \sqrt{l(l+1) + \frac{1}{4}} = \left(l + \frac{1}{2}\right)$. Here prime denotes derivative w. r. t. the radial coordinate r.

In the presence of an external field, one can perform a minimal substitution $\vec{p}^2 \rightarrow (\vec{p} - e\vec{A})^2$ into the Schrodinger wave equation [2, 3, 5, 76], where $\vec{A}$ is the electromagnetic three-vector potential given by

$$A_r = 0 = A_\theta \quad , \quad A_\phi = \frac{\Phi_{AB}}{2\pi r \sin\theta}. \tag{24}$$

Therefore, Eq. (15) can be written as

$$\left[\frac{1}{r^2}\frac{d}{dr}\left(r^2\frac{d}{dr}\right) + \frac{1}{r^2\sin\theta}\frac{d}{d\theta}\left(\sin\theta\frac{d}{d\theta}\right) + \frac{1}{r^2\sin^2\theta}\left(\frac{d}{d\phi} - i\alpha\right)^2 + \frac{2M}{\hbar^2}(E - V(r))\right]\Psi(r,\theta,\phi) = 0, \tag{25}$$

Where $\alpha = \frac{\Phi_{AB}}{\Phi_0}$ the amount of quantum flux and $\Phi_0 = 2\pi e^{-1}$ is the quantum flux. Here $\Phi_{AB} = const$ is the Aharonov-Bohm magnetic flux [2, 3, 5, 75, 76, 78-85]. Note that for the above mentioned flux field, there is no magnetic field $\vec{B} = \vec{\nabla} \times \vec{A} = 0$, that is, no direct interaction of the non-relativistic particle with the magnetic field but the particle is confined to the flux field.

Following the previous procedure, one can find the following radial wave equation

$$\Rightarrow U'' + \frac{1}{r}U' + \left[\frac{2M}{\hbar^2}(E - V) - \frac{J_0^2}{r^2}\right]U = 0, \tag{26}$$

where $J_0 = l_0 + \frac{1}{2}$, $l_0 = (l - \alpha)$, and replacing m by $m_0$ with $m_0 = (m - \alpha)$ and $\lambda_0 = l_0(l_0 + 1)$.

## 4. (A) Effects of Modified Coulomb-potential $V = a - \frac{b}{r}$.

In this section, we consider a potential which is the sum of a constant plus attractive Coulomb potential given in Eq. (3). This type of potential we called the modified Coulomb potential and have wide application in physics and chemistry. We use this type of potential in the above mentioned quantum mechanical problem and obtained the eigenvalue solution.

Therefore, substituting the modified Coulomb potential into the Eq. (A3), we have

$$U''(r) + \frac{1}{r}U'(r) + \left[\frac{2M}{\hbar^2}\left(E - a + \frac{b}{r}\right) - \frac{J_0^2}{r^2}\right]U(r) = 0$$

$$\Rightarrow U''(r) + \frac{1}{r}U'(r) + \left[\frac{2M}{\hbar^2}(E - a) + \frac{2Mb}{\hbar^2 r} - \frac{J_0^2}{r^2}\right]U(r) = 0$$

$$\Rightarrow U''(r) + \frac{1}{r}U'(r) + \left[\Delta + \frac{\omega^2}{r} - \frac{J_0^2}{r^2}\right]U(r) = 0. \quad (27)$$

Where,

$$\Delta = \frac{2M}{\hbar^2}(E - a), \quad \omega^2 = \frac{2Mb}{\hbar^2}. \quad (28)$$

Equation (24) can be written as

$$U''(r) + \frac{1}{r}U'(r) + \frac{1}{r^2}\left[\Delta r^2 + \omega^2 r - J_0^2\right]U(r) = 0. \quad (29)$$

Comparing equation (26) with Eq. (7), we have

$$\xi_1 = -\Delta, \qquad \xi_2 = \omega^2, \qquad \xi_3 = J_0^2,$$

$$\alpha_1 = 1, \quad \alpha_2 = 0, \quad \alpha_3 = 0, \quad \alpha_4 = 0, \quad \alpha_5 = 0, \quad \alpha_6 = \alpha_5^2 + \xi_1 = \xi_1,$$

$$\alpha_7 = 2\alpha_4\alpha_5 - \xi_2 = 2.0.0 - (\omega^2) = \omega^2, \quad \alpha_8 = \alpha_4^2 + \xi_3 = 0 + \xi_3 = J_0^2,$$

$$\alpha_9 = \alpha_6 + \alpha_3\alpha_7 + \alpha_3^2\alpha_8 = -\Delta + 0(\omega^2) + 0 = -\Delta,$$

$$\alpha_{10} = \alpha_1 + 2\alpha_4 + 2\sqrt{\alpha_8} = 1 + 2\times 0 + 2J_0 = 1 + 2J_0,$$

$$\alpha_{11} = \alpha_2 - 2\alpha_5 + 2(\sqrt{\alpha_9} + \alpha_3\sqrt{\alpha_8}) = 0 - 2\times 0 + 2(\sqrt{-\Delta} + 0\sqrt{J_0^2}) = 2\sqrt{-\Delta},$$

$$\alpha_{12} = \alpha_4 + \sqrt{\alpha_8} = 0 + \sqrt{J_0^2} = J_0,$$

$$\alpha_{13} = \alpha_5 - (\sqrt{\alpha_9} + \alpha_3\sqrt{\alpha_8}) = 0 - (\sqrt{-\Delta} + 0) = -\sqrt{-\Delta}. \quad (30)$$

Substituting Eq. (27) into the energy equation (8), we have

$$\Rightarrow (2n + 1)\sqrt{-\Delta} - \omega^2 + 2J_0\sqrt{-\Delta} = 0$$

$$\Rightarrow (2n + 1 + 2J_0)\sqrt{-\Delta} = \omega^2$$

$$\Rightarrow E_{n,l} = a - \frac{M b^2}{2\hbar^2 (n+1+l-\alpha)^2}, \tag{31}$$

where the radial quantum number n=0,1,2,3,…

Therefore the wave function becomes,

$$U(r) = r^{\alpha_{12}} e^{\alpha_{13}} L_n^{\alpha_{10}-1} \alpha_{11} r$$

$$\Rightarrow R_{n,l}(r) = r^{J_0 - \frac{1}{2}} \cdot e^{-\sqrt{-\Delta} r} L_n^{(1+2J_0)-1}(2\sqrt{-\Delta}\, r)$$

$$\Rightarrow R_{n,l}(r) = -r^{(l-\alpha)} e^{-\frac{M b r}{2\hbar^2 (n+1+l-\alpha)}} L_n^2 \left[\frac{M b}{\hbar^2 (n+1+l-\alpha)}\right]. \tag{32}$$

Equation (29) is the non-relativistic energy eigenvalue and Eq. (29) is the un-normalized radial wave function of a Schrodinger particle in the presence of Aharonov-Bohm flux field subject to the modified Coulomb potential.

For $a \to 0$, the modified potential becomes Coulomb potential. In that case, the energy expression Eq. (28) reduces to

$$E_{n,l} = -\frac{M b^2}{2\hbar^2 (n+1+l-\alpha)^2}. \tag{33}$$

Equation (30) is the modified eigenvalue expression of an electron in case of a hydrogen atom in the presence of Aharonov-Bohm flux field subject to an attractive Coulomb potential. Thus, we can see that the presence of a constant potential (V0=a) and magnetic flux modified the non-relativistic energy eigenvalue Eq. (31-28) of a Schrodinger particle in comparison to the case with a Coulomb potential.

## 4. (B) Effects of modified harmonic oscillator potential $V = a + b\, r^2$.

In this section, we study the non-relativistic particle under the influence of Aharonov-Bohm flux field with a physical potential equal to a constant plus harmonic oscillator potential called the modified harmonic oscillator potential. Substituting this type of potential into the Eq. (26-A3), we have

$$U''(r) + \frac{1}{r}U'(r) + \left[\frac{2M}{\hbar^2}(E - a - br^2) - \frac{J_0^2}{r^2}\right]U(r) = 0$$

$$\Rightarrow U''(r) + \frac{1}{r}U'(r) + \left[\frac{2M}{\hbar^2}(E - a) - \frac{2Mb}{\hbar^2}r^2 - \frac{J_0^2}{r^2}\right]U(r) = 0$$

$$\Rightarrow U''(r) + \frac{1}{r}U'(r) + \left[\nabla - \omega^2 r^2 - \frac{J_0^2}{r^2}\right]U(r) = 0, \qquad (34)$$

Where

$$\Delta = \frac{2ME}{\hbar^2}(E\text{-}a) \quad , \quad \omega^2 = \frac{2Mb}{\hbar^2}.$$

Now, we introduce a new variable via $s = \omega r^2$ into the Eq. (34-31) as follows:

$$U'(r) = U''(s)\, 2\,\omega\, r$$

$$\therefore U''(r) = 2\,\omega\, U'(s) + U''(s)(2\,\omega\, r)^2$$

$$\Rightarrow U''(r) = 2\,\omega\, U'(s) + 4\,\omega^2 r^2 U''(s)$$

$$\Rightarrow U''(r) = 2\,\omega\, U'(s) + 4\,\omega\, s\, U''(s). \qquad (35)$$

From equation (11), we have

$$2\omega\, U'(s) + 4\,\omega\, s\, U''(s) + 2\,\omega\, U'(s) + \left(\Delta - \omega s - \frac{J_0^2 \omega}{s}\right) U(s) = 0$$

$$\Rightarrow 4\,\omega\, s\, U''(s) + 4\,\omega\, U'(S) + \left(\Delta - \omega s - \frac{J_0^2 \omega}{s}\right) U(s) = 0$$

$$\Rightarrow U''(s) + \frac{1}{s} U'(s) + \frac{1}{4\omega s}\left(\Delta - \omega s - \frac{J_0^2 \omega}{s}\right) U(s) = 0$$

$$\Rightarrow U''(s) + \frac{1}{s} U'(s) + \frac{1}{s^2}\left(\frac{\Delta}{4\omega} s - \frac{1}{4} s^2 - \frac{J_0^2}{4}\right) U(s) = 0$$

$$\Rightarrow U''(s) + \frac{1}{s} U'(s) + \frac{1}{s^2}\left(-\xi_1 s^2 + \xi_2 s - \xi_3\right) U(s) = 0. \tag{36}$$

Where we have defined

$$\xi_1 = \frac{1}{4}, \qquad \xi_2 = \frac{\Delta}{4\omega}, \qquad \xi_3 = \frac{J_0^2}{4}. \tag{37}$$

Comparing Eq. (33) with Eq. (7), we have

$$\alpha_1 = 1, \quad \alpha_2 = 0, \quad \alpha_3 = 0, \quad \alpha_4 = \tfrac{1}{2}(1-\alpha_1) = 0, \quad \alpha_5 = \tfrac{1}{2}(\alpha_2 - \alpha_3) = 0,$$

$$\alpha_6 = \tfrac{1}{2}(\alpha_5^2 + \xi_1) = \tfrac{1}{2}\left(0 + \tfrac{1}{4}\right) = \tfrac{1}{4}, \quad \alpha_7 = 2\alpha_4 \alpha_5 - \xi_2 = 0 - \tfrac{\Delta}{4\omega} = -\tfrac{\Delta}{4\omega},$$

$$\alpha_8 = \alpha_4^2 + \xi_3 = \frac{J_0^2}{4},$$

$$\alpha_9 = \alpha_6 + \alpha_3 \alpha_7 + \alpha_3^2 \alpha_8 = 1/4,$$

$$\alpha_{10} = \alpha_1 + 2\alpha_4 + 2\sqrt{\alpha_8} = 1 + 0 + 2\sqrt{\tfrac{J_0^2}{4}} = 1 + J_0,$$

$$\alpha_{11} = \alpha_2 - 2\alpha_5 + 2(\sqrt{\alpha_9} + \alpha_3\sqrt{\alpha_8}) = 2\sqrt{\tfrac{1}{4}} = 1,$$

$$\alpha_{12} = \alpha_4 + \sqrt{\alpha_8} = \sqrt{\tfrac{J_0^2}{4}} = \tfrac{J_0}{2},$$

$$\alpha_{13} = \alpha_5 - \sqrt{\alpha_9} + \alpha_3\sqrt{\alpha_8} = 0 - \sqrt{\frac{1}{4}} = -\frac{1}{2}. \tag{38}$$

Substituting Eqs. (37-34)---(38-35) into the energy equation (8), we have

$$\Rightarrow \frac{1}{2}(2n+1) - \frac{\Delta}{4\omega} + \frac{J_0}{2} = 0$$

$$\Rightarrow \frac{\Delta}{2\omega} = 2n + 1 + J_0$$

$$\Rightarrow E_{n,l} = a + \hbar\sqrt{\frac{2b}{M}}\left[2n + \frac{3}{2} + l - \alpha\right]. \tag{39}$$

And the radial wave-function is given by

$$U(s) = s^{\alpha_{12}} e^{\alpha_{13} s} L_n^{(\alpha_{10}-1)}(\alpha_{11} s)$$

$$\Rightarrow U_{n,l}(s) = s\, e^{-\frac{s}{2}} L_n^{J_0}(s)$$

$$\Rightarrow U_{n,l}(r) = \omega\, r^2 e^{-\frac{1}{2}\omega r^2} L_n^{l_0 + \frac{1}{2}}(\omega r^2). \tag{40}$$

Equation (39-36) is the energy profile and Eq. (40-37) is the un-normalized radial wave-function of non-relativistic particle. For $a \to 0$, the eigenvalue expression Eq. (39-36) reduces to

$$E_{n,l} = \hbar\sqrt{\frac{2b}{M}}\left[2n + \frac{3}{2} + l - \alpha\right]. \tag{41}$$

The above energy eigenvalue is the modified energy profile of a linear harmonic oscillator in quantum mechanics in three-dimensional case given in many textbooks, where the oscillator frequency is defined by $\omega_o = \sqrt{\frac{2b}{M}}$. Thus, we can see that the presence of a constant potential ($V_0 = a$) in the harmonic oscillator potential and the Aharonov-Bohm flux field modified the energy eigenvalue and the wave-function of a harmonic oscillator.

## 4. (C) Effect of the Kratzer-Feus Potential (Inverse Square plus Coulomb Potential) $V(r) = -\frac{b}{r} + \frac{c}{r^2}$ [72-77].

In this section, we study the non-relativistic particle under the influence of Aharonov-Bohm flux field subject to a physical potential called the Kratzer-Feus potential (combination of Coulomb plus inverse square potentials). For $b \to 0$, we get back an inverse square potential while for $c \to 0$, one will have an attractive Coulomb potential.

Thereby, substituting this type of potential into the Eq. (26), we have

$$U''(r) + \frac{1}{r}U'(r) + \left[\frac{2M}{\hbar^2}\left(E + \frac{b}{r} - \frac{c}{r^2}\right) - \frac{J_0^2}{r^2}\right]U(r) = 0$$

$$\Rightarrow U''(r) + \frac{1}{r}U'(r) + \left[\frac{2M}{\hbar^2}E + \frac{2Mb}{\hbar^2 r} - \frac{2Mc}{\hbar^2 r^2} - \frac{J_0^2}{r^2}\right]U(r) = 0$$

$$\Rightarrow U''(r) + \frac{1}{r}U'(r) + \left[\frac{2ME}{\hbar^2} + \frac{2Mb}{\hbar^2}\frac{1}{r} - \left(\frac{2Mc}{\hbar^2} + J_0^2\right)\frac{1}{r^2}\right]U(r) = 0$$

$$\Rightarrow U''(r) + \frac{1}{r}U'(r) + \left[\Lambda + \frac{\omega^2}{r} - \frac{\sigma^2}{r^2}\right]U(r) = 0$$

$$\Rightarrow U''(r) + \frac{1}{r}U'(r) + \frac{1}{r^2}(\Lambda r^2 + \omega^2 r - \sigma^2)U = 0, \tag{42}$$

Where,

$$\Lambda = \frac{2ME}{\hbar^2} \quad , \quad \omega^2 = \frac{2Mb}{\hbar^2} \quad , \quad \sigma^2 = \frac{2Mc}{\hbar^2} + J_0^2. \tag{43}$$

Comparing Eq. (42-39) with Eq. (7), we have

$$\xi_1 = -\Lambda \quad , \quad \xi_2 = \omega^2 \quad , \quad \xi_3 = \sigma^2,$$

$$\alpha_1 = 1, \quad \alpha_2 = 0, \quad \alpha_3 = 0, \quad \alpha_4 = \frac{1}{2}(1 - \alpha_1) = 0,$$

$$\alpha_5 = \frac{1}{2}(\alpha_2 - 2\alpha_3) = 0, \quad \alpha_6 = \xi_1 = -\Lambda, \quad \alpha_7 = 2\alpha_4\alpha_5 - \xi_2 = 0 - \xi_2 = -\omega^2,$$

$$\alpha_8 = \alpha_4^2 + \xi_3 = 0 + \xi_3 = \xi_3 = \sigma^2,$$

$$\alpha_9 = \alpha_6 + \alpha_3\alpha_7 + \alpha_3^2\alpha_8 = \xi_1 + 0 + 0 = \xi_1 = -\Lambda,$$

$$\alpha_{10} = \alpha_1 + 2\alpha_4 + 2\sqrt{\alpha_8} = 1 + 0 + 2\sqrt{\xi_3} = 1 + 2\sqrt{\xi_3} = 1 + 2\sigma,$$

$$\alpha_{11} = \alpha_2 - 2\alpha_5 + 2(\sqrt{\alpha_9} + \alpha_3\sqrt{\alpha_8}) = 0 - 0 + 2(\sqrt{\xi_1} + 0) = 2\sqrt{\xi_1} = 2\sqrt{-\Lambda},$$

$$\alpha_{12} = \alpha_4 + \sqrt{\alpha_8} = 0 + \sqrt{\xi_3} = \sqrt{\xi_3} = \sigma,$$

$$\alpha_{13} = \alpha_5 - (\sqrt{\alpha_9} + \alpha_3\sqrt{\alpha_8}) = 0 - (\sqrt{\xi_1} + 0) = -\sqrt{\xi_1} = -\sqrt{-\Lambda}. \tag{44}$$

Substituting Eq. (44-41) into the energy equation (8), we have

$$\Rightarrow (2n+1)\sqrt{\xi_1} - \xi_2 + 2\sqrt{\xi_1\xi_3} = 0$$

$$\Rightarrow (2n+1)\sqrt{-\Lambda} - \omega^2 + 2\sqrt{-\Lambda\,\sigma^2} = 0$$

$$\Rightarrow E_{n,l} = -\frac{M\,b^2}{2\hbar^2\left[n+\frac{1}{2}+\sqrt{\frac{2\,M\,c}{\hbar^2}+\left(l-\alpha+\frac{1}{2}\right)^2}\right]^2}. \tag{45}$$

Therefore the wave-function becomes

$$U(r) = r^{\alpha_{12}} e^{\alpha_{13} r} L_n^{\alpha_{10}-1}(\alpha_{11} r)$$

$$\Rightarrow U_{n,l}(r) = r^\sigma e^{-\sqrt{-\nabla}r} L_n^{(1+2\sigma)-1}(2\sqrt{-\Lambda}\,r)$$

$$\Rightarrow U_{n,l}(r) = r^\sigma e^{-\frac{M\,b}{\hbar^2\left(n+\frac{1}{2}+\sigma\right)}} L_n^{(2\sigma)}\left[\frac{2\,M\,b}{\hbar^2\left(n+\frac{1}{2}+\sigma\right)}r\right]. \tag{46}$$

Since $\sqrt{-\Lambda} = \frac{M\,b}{\hbar^2\left(n+\frac{1}{2}+\sigma\right)} = \frac{M\,b}{\hbar^2\left(n+\frac{1}{2}+\sqrt{\frac{2\,M\,c}{\hbar^2}+\left(l-\alpha+\frac{1}{2}\right)^2}\right)}$ and $\sigma = \sqrt{\frac{2\,M\,c}{\hbar^2}+\left(l-\alpha+\frac{1}{2}\right)^2}$.

Equation (45) is the non-relativistic energy eigenvalue and Eq. (46) is the un-normalized radial wave function. For zero Aharonov-Bohm flux $\Phi_{AB} \to 0$, that is, $\alpha \to 0$, the energy profile Eq. (45) reduces to the results obtained in Ref. [72]. Thus, we can see that the presence the Aharonov-Bohm flux field modified the energy eigenvalue and the wave-function of a non-relativistic particle subject to the Kratzer-Feus potential.

## 4. (D) Effects of the Mie-type Potential (Constant plus Coulomb and Inverse square potentials) $V = a - \frac{b}{r} + \frac{c}{r^2}$ [20, 77].

In this section, we study the non-relativistic particle under the influence of Aharonov-Bohm flux field subject to the Mie-type potential (combination of a constant plus attractive Coulomb and inverse square potentials). For $a \to 0$ and b$\to$ 0, we will get back an inverse square potential. For $b \to 0$ and $c \to 0$, one will have a constant potential. For $a \to 0$ and c$\to$ 0, we get back a Coulomb potential. In addition, for $a \to 0$, we have the Kratzer-Feus potential which we discussed in the previous section.

Thereby, substituting this type of potential into the Eq. (26), we have

$$U''(r) + \frac{1}{r}U'(r) + \left[\frac{2M}{\hbar^2}\left(E - a + \frac{b}{r} - \frac{c}{r^2}\right) - \frac{J_0^{\,2}}{r^2}\right]U(r) = 0$$

$$U''(r) + \frac{1}{r}U'(r) + \left[\frac{2M}{\hbar^2}(E - a) + \frac{2M\,b}{\hbar^2\,r} - \frac{2M\,c}{\hbar^2\,r^2} - \frac{J_0^{\,2}}{r^2}\right]U(r) = 0$$

$$U''(r) + \frac{1}{r}U'(r) + \left[\Delta + \frac{\omega^2}{r} - \frac{\sigma^2}{r^2}\right]U(r) = 0, \tag{47}$$

Where $\Delta = \frac{2M}{\hbar^2}(E-a)$, $\omega^2 = \frac{2M b}{\hbar^2}$, $\sigma^2 = \frac{2M c}{\hbar^2} + J_0^2$.

Equation (44) can be expressed as

$$U''(r) + \frac{1}{r}U'(r) + \frac{1}{r^2}[\Delta r^2 + \omega^2 r - \sigma^2]U(r) = 0. \tag{48}$$

Comparing equation (48) with Eq. (7), we have

$$\xi_1 = -\Delta, \qquad \xi_2 = \omega^2, \qquad \xi_3 = \sigma^2,$$

$$\alpha_1 = 1, \quad \alpha_2 = 0, \quad \alpha_3 = 0, \quad \alpha_4 = \frac{1}{2}(1-\alpha_1) = 0, \quad \alpha_5 = \frac{1}{2}(\alpha_2 - 2\alpha_3) = 0$$

$$\alpha_6 = \alpha_5^2 + \xi_1 = 0 + \xi_1 = \xi_1,$$

$$\alpha_7 = 2\alpha_4\alpha_5 - \xi_2 = 2\times 0 \times 0 - (\omega^2) = -\omega^2,$$

$$\alpha_8 = \alpha_4^2 + \xi_3 = 0 + \xi_3 = \sigma^2,$$

$$\alpha_9 = \alpha_6 + \alpha_3\alpha_7 + \alpha_3^2\alpha_8 = -\Delta + 0(-\omega^2) + 0 = -\Delta,$$

$$\alpha_{10} = \alpha_1 + 2\alpha_4 + 2\sqrt{\alpha_8} = 1 + 2\sigma,$$

$$\alpha_{11} = \alpha_2 - 2\alpha_5 + 2(\sqrt{\alpha_9} + \alpha_3\sqrt{\alpha_8}) = 0 - 2\times 0 + 2(\sqrt{-\Delta} + 0\sqrt{\sigma^2}) = 2\sqrt{-\Delta},$$

$$\alpha_{12} = \alpha_4 + \sqrt{\alpha_8} = 0 + \sqrt{\sigma^2} = \sigma,$$

$$\alpha_{13} = \alpha_5 - (\sqrt{\alpha_9} + \alpha_3\sqrt{\alpha_8}) = 0 - (\sqrt{-\Delta} + 0) = -\sqrt{-\Delta}. \tag{49}$$

Substituting Eq. (49) into the Eq. (8), we have

$$\Rightarrow (2n+1)\sqrt{-\Delta} - \omega^2 + 2\sigma\sqrt{-\Delta} = 0$$

$$\Rightarrow E_{n,l} = a - \frac{M b^2}{2\hbar^2\left(n + \frac{1}{2} + \sqrt{\frac{2M c}{\hbar^2} + \left(l - \alpha + \frac{1}{2}\right)^2}\right)^2}, \tag{50}$$

where n=0,1,2,3,......

Therefore the wave function becomes,

$$U(r) = r^{\alpha_{12}} e^{\alpha_{13}} L_n^{\alpha_{10}-1}(\alpha_{11} r)$$

$$U_{n,l}(r) = r^{\sigma} \cdot e^{-\sqrt{-\Delta}\, r} L_n^{(1+2\sigma)-1}(2\sqrt{-\Delta}\, r)$$

$$U_{n,l}(r) = r^{\sigma}\, e^{-\frac{M b}{2\hbar^2 (n+\frac{1}{2}+\sigma)}}\, L_n^{(2\sigma)}\left[\frac{M b}{\hbar^2 (n+\frac{1}{2}+\sigma)} r\right], \qquad (51)$$

$$\sigma = \sqrt{\frac{2 M c}{\hbar^2} + \left(l - \alpha + \frac{1}{2}\right)^2}.$$

Equation (50) is the energy eigenvalue and Eq. (51) is the un-normalized radial wave function of a non-relativistic particle in the presence of Aharonov-Bohm flux field subject to the Mie-type potential. For zero Aharonov-Bohm flux $\Phi_{AB} \to 0$, that is, $\alpha \to 0$, the energy profile Eq. (50) reduces to the results obtained in Ref. [72]. Thus, we can see that the presence Aharonov-Bohm flux field modified the energy eigenvalue and the wave-function of a non-relativistic particle subject to the Mie-type potential.

## 5. Conclusions:

The spectral problem of the Schrödinger equation with spherically symmetric potential is important in the spectroscopy of complex chemical compounds and molecules. It is also important in describing the spectra of hadrons reasons spherically mesons Quarkoniums system. It is know that despite being a rigorous theoretical approach, potential modes gives us a satisfactory description of the mass spectra for such system as quarkonium. The interaction in such system is usually represented by confining type potential. As an example one can take the sum of Carnell and inverse quadratic potential consisting of three terms, one of the terms is responsible for the coulomb interaction of quarks. Second to the string interaction and third terms corresponds inverse quadratic potential.

In this project, we have derived the radial wave equation of the Schrödinger equation for a spherically symmetric potential in the presence of Aharonov-Bohm flux field. Then, we have solved this radial wave equation under the influence of some physical potential using the Nikiforov-Uvorov method and obtained the eigenvalue solutions. We have shown that for some physical potential, the eiegnvalue solution of the non-relativistic particle gets modified due to the presence of Aharonov-Bohm flux field in comparison to the known results obtained in the literature. It is well-known that the dependence of the eigenvalue solutions on the geometric quantum phase gives us an analogue of the electromagnetic Aharonov-Bohm effect for bound states [78-85].

In *sub-section 4(A)*, we have considered a modified Coulomb potential of the form $V(r) = \left(a - \frac{b}{r}\right)$ and solved the radial Schrödinger equation using the NU-method. The energy profile given by the Eq. (28) and the un-normalized radial wave-function given by the Eq. (19) are obtained. We have seen that due to the presence of a constant potential term in Coulomb potential and quantum flux modified the energy spectrum of the non-relativistic particle.

In s*ub-section 4(B)*, we have considered a modified harmonic oscillator potential $V(r) = (a + b\, r^2)$ and solved the radial Schrödinger equation using the NU-method. The energy profile given by the Eq. (36) and the un-normalized radial wave-function given by the Eq. (37) are obtained. We have seen that due to the presence of a constant potential with the harmonic oscillator potential and the quantum flux modified the energy spectrum of the non-relativistic particle in comparison to the known results for a linear harmonic oscillator potential in three dimensional cases.

In s*ub-section 4(C),* we have considered the Kratzer-Feus potential (Coulomb plus inverse square potential) $V(r) = \left(-\frac{b}{r} + \frac{c}{r^2}\right)$ [72---77] and have solved the radial Schrödinger wave equation using the NU-method. The energy profile given by the Eq. (42) and the un-normalized radial wave-function given by the Eq. (43) are obtained. We have seen that the energy spectrum of the non-relativistic particle gets modified due to the presence of Aharonov-Bohm flux field in comparison to the result known in the literature for Kratzer-Feus potential.

In s*ub-section 4(D),* we have considered the Mie-type potential (constant plus Coulomb and inverse square potentials) $V(r) = \left(a - \frac{b}{r} + \frac{c}{r^2}\right)$ [20, 77] in the quantum system and solved the radial Schrödinger equation. The energy spectrum given by the Eq. (47) and the un-normalized radial wave-function given by the Eq. (48) are obtained. We have seen that the energy spectrum of the non-relativistic particle gets modified due to the presence of Aharonov-Bohm flux field in comparison to the known results obtained in the literature.

We can conclude that our presented results in this study are expected to enable possibilities for pure theoretical and experimental physicists because the results are exact and more general. In each case, we have seen in the energy eigenvalue expression that the angular quantum number *l* is shifted, $l \to l_0 = \left(l - \frac{e\,\Phi_{AB}}{2\,\pi}\right)$, an effective angular momentum quantum number. Thus, we have that $E_{n,l}(\Phi_{AB} \pm \Phi_0 \vartheta) = E_{n,l \mp \vartheta}(\Phi_{AB})$, that is, the energy eigenvalue is a periodic function of the geometric quantum phase which gives us an analogue of the electromagnetic Aharonov-Bohm effect for bound-states [78-85].

**Acknowledgment:** We sincerely thank anonymous kind referee(s) for his/her valuable comments and helpful suggestions. This work is a part of our Master of Science (MSc) Project dissertation having numbers **2020/MSP/0011, 0017, 0026 & 0034**.

**Contribution:** **F. Ahmed-**Conceptualization, Methodology, Validation, Review & Editing, Supervision; **K. Ahmed**-section 4(A) & Writing; **A. Ahmed**-section 4(B); **A. Islam**- section 4 (C); **B. P. Barman**-section 4(D).

**Data Availability:** No new data are generated in this project.

**Funding Statement:** No fund has received for this paper.